\documentclass[12pt]{article}

\begin{document}

\newcommand{\siml}{\stackrel{<}{\sim}}
\newcommand{\simg}{\stackrel{>}{\sim}}
\newcommand{\lleq}{\stackrel{<}{=}}



%
\begin{center}
{\large\bf
Generalized Rate-Code Model for Neuron Ensembles \\
with Finite Populations
} 
\end{center}

\begin{center}
Hideo Hasegawa
\footnote{Electronic address:  hasegawa@u-gakugei.ac.jp}
\end{center}

\begin{center}
{\it Department of Physics, Tokyo Gakugei University  \\
Koganei, Tokyo 184-8501, Japan}
\end{center}
\begin{center}
({\today})
\end{center}
\thispagestyle{myheadings}

\begin{abstract}
We have proposed a generalized Langevin-type rate-code model
subjected to multiplicative noise, 
in order to study stationary and dynamical properties 
of an ensemble containing {\it finite} $N$ neurons.
Calculations using the Fokker-Planck equation (FPE)  
have shown that owing to the multiplicative noise,
our rate model yields various 
kinds of stationary non-Gaussian distributions 
such as gamma, 
inverse-Gaussian-like and log-normal-like distributions,
which have been experimentally observed.
Dynamical properties of the rate model have been studied
with the use of the augmented moment method (AMM),
which was previously proposed by the author 
with a macroscopic point of view
for finite-unit stochastic systems.
In the AMM, 
original $N$-dimensional stochastic
differential equations (DEs) are
transformed into three-dimensional deterministic
DEs for means and fluctuations of local and global variables.
Dynamical responses of the neuron ensemble to pulse and sinusoidal
inputs calculated by the AMM are in good agreement
with those obtained by direct simulation.
The synchronization in the neuronal ensemble is discussed.
Variabilities of the firing rate
and of the interspike interval (ISI) are shown to increase
with increasing the magnitude of multiplicative noise,
which may be a conceivable origin of the observed large
variability in cortical neurons.
\end{abstract}

\noindent
\vspace{0.5cm}

%

\vspace{2cm}

\newpage

\section{Introduction}

Neurons in a brain communicate information,
emitting spikes which propagate through axons and
dendrites to neurons at the next stage.
It has been a long-standing controversy whether
information of neurons is encoded in the 
firing rates ({\it rate code}) or in
the more precise firing times ({\it temporal code})
\cite{Rieke96,Ursey99,deCharms00}.
Some experimental results having been reported
seem to support the
former code while some the latter
\cite{Rieke96,Ursey99,deCharms00}.
In particular,
a recent success in brain-machine interface (BMI) 
\cite{Anderson04}\cite{Chapin99} 
suggests that the population rate code is employed in
sensory and motor neurons while it is still not clear
which code is adopted in higher-level cortical neurons.

Experimental observations have shown that
in many areas of the brain, neurons are organized into groups of cells
such as columns in the visual cortex \cite{Mount57}.
A small patch in cortex 
contains thousands of similar neurons, which receive inputs
from the same patch and other patches.
There are many theoretical studies on the property of 
neuronal ensembles consisting of equivalent neurons,
with the use of spiking neuron models 
or rate-code models 
(for a review on neuronal models, 
see \cite{Gerstner02}; related references therein).
In the spiking neuron model, the dynamics of the membrane
potential of a neuron in the ensemble is described 
by the Hodgkin-Huxley (HH)-type nonlinear differential equations (DEs)
\cite{Hodgkin52}
which express the conductance-based mechanism for firings.
Reduced, simplified models such as the integrate-and-fire (IF)
and FitzHugh-Nagumo (FN) models have been also employed.
In contrast, in the rate-code model, neurons are
regarded as transducers between input and output signals,
both of which are expressed in terms of spiking rates. 

Computational neuroscientists have tried to understand the 
property of ensemble neurons by using the two approaches:
direct simulations (DSs) and analytical approaches. 
DS calculations have been performed for large-scale networks mostly
described by the simplest IF model.
Since the computational time of DS grows as $N^2$ with $N$,
the size of the ensemble, a large-scale
DS with more realistic models becomes difficult.
Although DS calculations provide us with useful insight
to the firing activity of the ensemble, it is desirable
to have results obtained by using analytical approaches.

Analytical or semi-analytical
calculation methods for neuronal ensembles have been proposed
by using the mean-field (MF) method \cite{Abbott93,Treves93, Gerstner95},
the population-density approaches \cite{Wilson72}-\cite{Eggert00},
the moment method \cite{Rod96}
and the augmented moment method (AMM) \cite{Hasegawa03a}
(details of the AMM will be discussed shortly).
It is interesting to analytically obtain the information of
the firing rate or the interspike interval (ISI), starting 
from the spiking neuron model.
It has been shown that the dynamics of the spiking neuron ensemble 
may be described by DEs of a macroscopic variable 
for the population density or
spike activity, which determines the firing rate of ensemble neurons
\cite{Wilson72}-\cite{Eggert00}.
By using the $f$-$I$ relation between the applied dc current $I$
and the frequency $f$ of autonomous firings, the rate-code model
for conduction-based neuron models is derived 
\cite{Amit91,Ermen94,Shriki03}.
When we apply the Fokker-Planck equation (FPE) method
to the neuron ensemble described by the IF model,
the averaged firing rate $R(t)$ is expressed by
$P(V, \theta, t)$
which denotes the distribution probability 
of the averaged membrane potential $V$ with
the threshold $\theta$ for the firing 
\cite{Brunel00}.

It is well known that neurons in brains are subjected to
various kinds of noise, though their precise origins
are not well understood. 
The response of neurons to stimuli is 
expected to be modified by noise in various ways.
Indeed, although firings of a single {\it in vitro} neuron 
are reported to be
precise and reliable \cite{Mainen95}, 
those of {\it in vivo} neurons are quite
unreliable due to noisy environment.
The strong criticism against the temporal code is that 
it is not robust against noise, while the rate code
is robust.

It is commonly assumed that there are two types of noise:
additive and multiplicative noise.
The magnitude of the former is independent of the
state of variable while that of the latter
depends on its state.
Interesting phenomena caused by the two noise have
been investigated \cite{Munoz04}.
It has been found that
the property of multiplicative noise
is different from that of additive noise
in some respects.
(1) Multiplicative noise induces a phase transition,
creating an ordered state, while additive noise
works to destroy the ordering \cite{Munoz05}\cite{Broeck94}.
(2) Although the probability distribution in stochastic systems
subjected to additive white noise follows a Gaussian,
multiplicative white noise generally yields a non-Gaussian 
distribution \cite{Tsallis88}-\cite{Hasegawa05b}.
(3) The scaling relation of the effective 
strength for additive noise given by
$\beta(N)=\beta(1)/\sqrt{N}$ is not applicable to
that for multiplicative noise:
$\alpha(N) \neq \alpha(1)/\sqrt{N}$, where $\alpha(N)$ and $\beta(N)$
denote effective strengths of multiplicative
and additive noise, respectively, in the $N$-unit system
\cite{Hasegawa06a}.
A naive approximation
of the scaling relation for multiplicative noise: 
$\alpha(N)=\alpha(1)/\sqrt{N}$
as adopted in Ref. \cite{Munoz05}, yields the result
which does not agree with that of DS
\cite{Hasegawa06a}.

In this paper, we will study the property of neuronal ensembles
based on the rate-code hypothesis.
Rate models having been proposed so far, are 
mainly given by \cite{Anderson04}
\begin{eqnarray}
\frac{dr_{i}(t)}{dt} &=& - \lambda r_{i}(t) 
+H\left(\frac{1}{N} \sum_j w_{ij}\:r_j(t)+I_i(t) \right)
+ \beta \xi_i(t), 
\end{eqnarray}
where 
$r_{i}(t)$ ($\geq 0$) denotes a firing rate of a neuron $i$ 
($i=1$ to $N$), 
$\lambda$ the relaxation rate,
$w_{ij}$ the coupling strength,
$H(x)$ the gain function,
$I_i(t)$ an external input,
and $\beta$ expresses the magnitude of additive white noise 
of $\xi_i(t)$ with the correlation:
$<\xi_{i}(t)\:\xi_{j}(t')> = \delta_{ij} \delta(t-t')$.
The rate model as given by Eq. (1) has been adopted
in many models based on neuronal population dynamics.
The typical rate model is the Wilson-Cowan model,
with which the stability of an ensemble consisting
of excitatory and inhibitory neurons is 
investigated \cite{Wilson72}\cite{Amari72}.
The rate model given by Eq. (1) with $H(x)=x$
is the Hopfield model \cite{Hopfield84},
which has been extensively adopted 
for studies on the memory in the brain
incorporating the plasticity of synapses into $w_{ij}$.
DS calculations have been performed,
for example, for a study of the population coding 
for $N=100$ \cite{Anderson04}.
Analytical studies of Eq. (1) are conventionally
made for the case of $N=\infty$,
adopting the FPE method with MF and diffusion approximations. 
The stationary distribution obtained by the FPE 
for Eq. (1) generally follows the Gaussian distribution.

ISI data obtained from 
experiments have been fitted by a superposition of
some known probability densities such as
the gamma, inverse-Gaussian and log-normal distributions
\cite{Gerstein64}-\cite{Vog05}.
The gamma distribution 
with parameters $\lambda$ and $\mu$
is given by
\begin{equation}
P_{gam}(x)=\frac{\mu^{-\lambda}}{\Gamma(\lambda)}\:x^{\lambda-1}
\exp\left(-{\frac{x}{\mu}} \right),
\end{equation}
which is derived from a simple 
stochastic IF model 
with additive noise
for Poisson inputs \cite{Tuckwell88},
$\Gamma(x)$ being the gamma function.
For $\lambda=1$ in Eq. (2), we get
the exponential distribution
describing a standard Poisson process.
The inverse Gaussian distribution 
with parameters $\lambda$ and $\mu$ given by
\begin{equation}
P_{IG}(x)=\left( \frac{\lambda}{2 \pi x^3} \right)^{1/2}
\exp \left[ -\frac{\lambda (x-\mu)^2}{2 \mu^2 x} \right],
\end{equation}
is obtained
from a stochastic IF model in which
the membrane potential is represented as a random walk
with drift \cite{Gerstein64}.  
The log-normal distribution 
with parameters $\mu$ and $\sigma$ given by
\begin{equation}
P_{LN}(x)=\frac{1}{\sqrt{2 \pi \sigma^2 }\:x}
\exp \left[- \frac{(\log x -\mu)^2}{2 \sigma^2}  \right],
\end{equation}
is adopted when
the log of ISI is assumed to follow a Gaussian form
\cite{McKeegan02}.
Fittings of experimental ISI data to a superposition
of these probability densities have been extensively
discussed in the literature
\cite{Gerstein64}-\cite{Vog05}.

The purpose of the present paper is to propose and
study the generalized, phenomenological 
rate model [Eqs. (5) and (6)].
We will discuss ensembles with {\it finite} populations,
contrary to most existing analytical theories 
except some ones ({\it e.g.} Ref. \cite{Eggert00}),
which discuss ensembles with {\it infinite} $N$. 
The stationary distribution of our rate model
will be discussed by using the FPE method.
It is shown that owing to the introduced multiplicative noise,
our rate model yields not only
the Gaussian distribution but also non-Gaussian distributions
such as gamma, inverse-Gaussian-like and log-normal-like
distributions.

The dynamical properties of our rate model will be studied by using
the augmented moment method (AMM) which was previously proposed 
by the present author
\cite{Hasegawa03a,Hasegawa06a,Hasegawa07a}.
Based on a macroscopic point of view,
Hasegawa \cite{Hasegawa03a} has
proposed the AMM,
which emphasizes not the property of individual neurons 
but rather that of ensemble neurons.
In the AMM,
the state of finite $N$-unit stochastic ensembles 
is described by a fairly small number of variables:
averages and fluctuations of local and global variables.
For example, the number of deterministic equation in the AMM 
becomes {\it three} for a $N$-unit Langevin model. 
The AMM has been successfully
applied to a study on the dynamics of 
the Langevin model and stochastic spiking neuron models
such as FN and HH models, 
with global, local 
or small-world couplings
(with and without transmission delays)
\cite{Hasegawa03b}-\cite{Hasegawa05a}. 

The AMM in \cite{Hasegawa03a} was originally
developed by expanding variables
around their stable mean values in order to obtain
the second-order moments both for
local and global variables in stochastic systems.  
In recent papers \cite{Hasegawa06a,Hasegawa07a},  
we have reformulated the AMM with the use of the FPE
to discuss stochastic systems subjected to multiplicative noise:
the FPE is adopted to avoid the difficulty
due to the Ito versus Stratonovich calculus inherent
to multiplicative noise. 
In the present paper, a study on the Langevin model 
with multiplicative noise made in \cite{Hasegawa07a}, 
has been applied to an investigation 
on the firing properties of neuronal ensembles.
Our method aims at the same purpose to effectively
study the property of neuronal ensembles as the approaches developed
in Refs. \cite{Wilson72}-\cite{Eggert00}\cite{Amit91}-\cite{Shriki03}.

The paper is organized as follows.
In Sec. 2, we discuss the generalized rate model for 
an ensemble containing $N$ neurons,
investigating its stationary and
dynamical properties. Some discussions are presented 
in Sec. 3, where variabilities of the rate and ISI
are calculated.  
The final Sec. 4 is devoted to our conclusion. 

\section{Property of neuron ensembles}

\subsection{Generalized rate-code model}

For a study of the property of a neuron ensemble
containing finite $N$ neurons,
we have assumed that the dynamics
of the firing rate $r_i(t)$ ($\geq 0$) of a neuron 
$i$ ($i=1$ to $N$) is given by 
\begin{eqnarray}
\frac{dr_{i}}{dt} &=& F(r_{i}) 
+H(u_{i})
+ \alpha G(r_{i}) \eta_{i}(t) + \beta \xi_{i}(t),
\end{eqnarray}
with
\begin{eqnarray}
u_{i}(t) &=& \left( \frac{w}{Z} \right) 
\sum_{j (\neq i)} \:r_{j}(t) + I_i(t).
\end{eqnarray}
Here
$F(x)$, $G(x)$ and $H(x)$ are arbitrary functions of $x$,
$Z$ $(=N-1)$ denotes the coordination number, 
$I_i(t)$ an input from external sources
and $w$ the coupling strength:
$\alpha$ and $\beta$ express the strengths of additive and
multiplicative noise, respectively, given by 
$\xi_i(t)$ and $\eta_i(t)$ expressing zero-mean Gaussian white
noise with correlations given by
\begin{eqnarray}
<\eta_{i}(t)\:\eta_{j}(t')> &=& \delta_{ij} \delta(t-t'), \\
<\xi_{i}(t)\:\xi_{j}(t')> &=& \delta_{ij} \delta(t-t'), \\
<\eta_{i}(t)\:\xi_{j}(t')> &=& 0.
\end{eqnarray}
The rate model in Eq. (1) adopts
$F(x)=-\lambda x$ and $G(x)=0$ (no multiplicative noise).

The gain function $H(x)$ expresses
the response of the firing rate ($r_i$) to
a synaptic input field ($u_i$).
It has been theoretically shown in \cite{Hasegawa00a} 
that when spike inputs with the mean ISI 
($T_{in}$) are applied to
an HH neuron, the mean ISI of output signals ($T_{out}$) 
is $T_{out} = T_{in}$ for $T_{in} \simg 15 $ ms
and $T_{out} \sim 15$ ms for $T_{in} \siml 15 $ ms.
This is consistent with the recent 
calculation for HH neuron
multilayers \cite{Wang06}, which shows a nearly linear
relationship between the input ($r_{in}$) 
and output rates ($r_{out}$) 
for $r_{in} < 60$ Hz  (Fig. 3 of Ref. \cite{Wang06}).
It is interesting that
the $r_{in}\:$-$\:r_{out}$ relation is continuous
despite the fact that the $f$-$I$ relation of
the HH neuron shows 
a discontinuous, type-II behavior according to Ref. \cite{Hodgkin52}.
In the literature,
two types of expressions for $H(x)$ have been adopted so far.
In the first category, sigmoid functions such as
$H(x)=1/(1+{\rm e}^{-x})$ ({\it e.g.} \cite{Wilson72}) and
${\rm arctan}(x)$ ({\it e.g.} \cite{Hayashi94})
have been adopted.
In the second category, gain functions such as
$H(x)=(x-x_c) \Theta(x-x_c)$ ({\it e.g.} \cite{Shriki03})
have been employed, 
modeling the $f$-$I$ function for the frequency $f$ 
of autonomous oscillation
against the applied dc current $I$,
$x_c$ expressing the critical value
and $\Theta(x)$ the Heaviside function:
$\Theta(x)=1$ for $x \geq 0$ and 0 otherwise.
The nonlinear, saturating behavior in $H(x)$ 
arises from the property of the refractory period
($\tau_r$) because spike outputs are prevented for 
$t_f < t < t_f+\tau_r$ after firing at $t=t_f$.
In this paper, we have adopted
a simple expression given by \cite{Monteiro02}
\begin{eqnarray}
H(x) &=& \frac{x}{\sqrt{x^2+1}},
\end{eqnarray}
although our results to be presented in the following sections,
are expected to be valid for any choice of $H(x)$.

\subsection{Stationary properties}
\subsubsection{Distribution of $r$}

By employing the FPE, we may discuss
the stationary distribution $p(r)$ for $w=0$
and $I_i(t)=I$, which is
given by \cite{Sakaguchi01,Anten02}
\begin{eqnarray}
{\rm ln} \:p(r) &\propto& X(r)+Y(r)
-\left( 1-\frac{\phi}{2} \right){\rm ln} 
\left[\frac{\alpha^2 G(r)^2}{2}+\frac{\beta^2}{2} \right],
\end{eqnarray}
with
\begin{eqnarray}
X(r) &=& 2 \int \:dr \:
\left[ \frac{F(r)}{\alpha^2 G(r)^2+\beta^2} \right], \\
Y(r) &=& 2 \int \:dr \:
\left[ \frac{H(I)}{\alpha^2 G(r)^2+\beta^2} \right],
\end{eqnarray}
where $\phi=0$ and 1 for Ito and Stratonovich
representations, respectively.
Hereafter we mainly adopt the Stratonovich representation.

\vspace{0.5cm}

\noindent
{\bf Case I} $F(x)=-\lambda x$ and $G(x)=x$

For the linear Langevin model, we get
\begin{eqnarray}
p(r) &\propto& 
\left[1+\left( \frac{\alpha^2 r^2}{\beta^2} \right)
\right]^{-(\lambda/\alpha^2+1/2)}
e^{Y(r)}, 
\end{eqnarray}
with
\begin{eqnarray}
Y(r)=\left( \frac{2 H}{\alpha \beta} \right)
{\rm arctan}\left( \frac{\alpha r}{\beta} \right),
\end{eqnarray}
where $H=H(I)$.
In the case of $H=Y(r)=0$, we get
the $q$-Gaussian given by
\cite{Sakaguchi01,Anten02}
\begin{eqnarray}
p(r) &\propto& \left[ 1-(1-q)\gamma r^2 \right]^{\frac{1}{1-q}},
\end{eqnarray}
with
\begin{eqnarray}
\gamma&=& \frac{2 \lambda+\alpha^2}{2 \beta^2}, \\
q&=& \frac{2 \lambda+3 \alpha^2}{2 \lambda +\alpha^2}.
\end{eqnarray}
We examine some limiting cases of Eq. (14) as follows.

\noindent
(a) For $\alpha=0$ and $\beta \neq 0$ 
({\it i.e.} additive noise only), 
Eq. (14) yields
\begin{eqnarray}
p(r) &\propto& e^{-\frac{\lambda}{\beta^2}(r- H/\lambda)^2}.
\end{eqnarray}

\noindent
(b) For $\beta=0$ and $\alpha \neq 0$ 
({\it i.e.} multiplicative noise only),
Eq. (14) leads to
\begin{eqnarray}
p(r) &\propto& r^{-(2\lambda/ \alpha^2+1)}
e^{-(2H/\alpha^2)/r}.
\end{eqnarray}

Distributions $p(r)$ calculated with the use of Eqs. (14)-(20)
are plotted in Figs. 1(a)-1(c). 
The distribution $p(r)$ for $\alpha=0.0$ 
(without multiplicative noise) in Fig. 1(a)
shows the Gaussian distribution
which is shifted by an applied input $I=0.1$.
When multiplicative noise is added ($\alpha \neq 0$), 
the form of $p(r)$ is changed to the $q$-Gaussian
given by Eq. (16). 
Fig. 1(b) shows that when the magnitude of additive noise
$\beta$ is increased,
the width of $p(r)$ is increased.
Fig. 1(c) shows that
when the magnitude of external input $I$ is increased, 
$p(r)$ is much shifted and widely spread. 
Note that for $\alpha=0.0$ and $\beta \neq 0$ (additive noise only),
$p(r)$ is simply shifted without a change
in its shape when increasing $I$ [Eq. (19)].

\vspace{0.5cm}
\noindent
{\bf Case II} $F(x)=-\lambda x^a$ and $G(x)=x^b$ ($a, b \ge 0$)

The special case of $a=1$ and $b=1$
has been discussed in the preceding case I [Eqs. (14)-(20)].
For arbitrary $a$ ($\ge 0$) and $b$ ($\ge 0$),
the probability distribution $p(r)$ given by Eqs. (11)-(13) becomes
\begin{eqnarray}
p(r) \propto 
\left[ 1+ \left( \frac{\alpha^2}{\beta^2}\right) 
\:r^{2b} \right]^{-1/2}\:e^{X(r)+Y(r)}, 
\end{eqnarray}
with
\begin{eqnarray}
X(r)&=& -\left( \frac{2 \lambda r^{a+1}}{\beta^2 (a+1)} \right)
F\left(1,\frac{a+1}{2 b}, \frac{a+1}{2b}+1
;-\frac{\alpha^2r^{2b}}{\beta^{2}} \right), \\
Y(r)&=& \left( \frac{2 H r}{\beta^2} \right)
F\left(1,\frac{1}{2 b}, \frac{1}{2b}+1
;-\frac{\alpha^2r^{2b}}{\beta^{2}} \right), 
\end{eqnarray}
where $F(a,b,c;z)$ is the hypergeometric function.
Some limiting cases of Eqs. (21)-(23) are shown in the following.

\noindent
(a) The case of $H=Y(r)=0$ was previously studied 
in \cite{Anten02}.

\noindent
(b) For $\alpha=0$ and $\beta \neq0$ ({\it i.e.} additive noise only), 
we get
\begin{eqnarray}
p(r) &\propto& 
\exp\left[ -\left(\frac{2\lambda}{\beta^2(a+1)} \right)r^{a+1}
+\left( \frac{2 H}{\beta^2} \right) r \right]. 
\end{eqnarray}

\noindent
(c) For $\beta=0$ and $\alpha \neq0$ 
({\it i.e.} multiplicative noise only), we get
\begin{eqnarray}
p(r) &\propto& r^{-b}
\exp\left[ -\left(\frac{2\lambda}{\alpha^2(a-2b+1)} \right) r^{a-2b+1}
-\left( \frac{2H}{\alpha^2(2b-1)} \right) r^{-2b+1} \right],
\nonumber \\
&&\hspace{3cm}\mbox{for $a-2b+1 \neq 0, 2b-1 \neq 0$} \\
&\propto& r^{-(2\lambda/\alpha^2+b)} 
\exp\left[ -\left (\frac{2H}{\alpha^2(2b-1)} \right) r^{-2b+1} \right],
\hspace{0.2cm}\mbox{for $a-2b+1 = 0$} \\
&\propto& r^{(2H/\alpha^2-1/2)} 
\exp\left[ -\left( \frac{2\lambda}{\alpha^2 a} \right) r^a \right], 
\hspace{1.5cm}\mbox{for $2b-1 = 0$} \\
&\propto& r^{-[2(\lambda-H)/\alpha^2+1/2]}, 
\hspace{2cm}\mbox{for $a-2b+1 = 0, 2b-1 = 0$}
\end{eqnarray}

\noindent
(d) In the case of $a=1$ and $b=1/2$, we get
\begin{eqnarray}
p(r) &\propto& \left( r+\frac{\beta^2}
{\alpha^2} \right)^{(2\lambda \beta^2/\alpha^4+2 H/\alpha^2-1/2)}
\exp\left[ -\left( \frac{2 \lambda}{\alpha^2} \right) r \right],
\end{eqnarray}
which reduces, in the limit of $\alpha=0$, to
\begin{eqnarray}
p(r) &\propto& \exp\left[-\left( \frac{\lambda}{\beta^2}\right)
\left(r-\frac{H}{\lambda} \right)^2 \right], 
\hspace{1cm}\mbox{for $\alpha=0$}
\end{eqnarray}

\vspace{0.5cm}

\noindent
{\bf Case III} $F(x)=-\lambda \ln x$ and $G(x)=x^{1/2}$

We get
\begin{eqnarray}
p(r) &\propto& r^{-1/2} 
\exp\left[ -\left( \frac{\lambda}{\alpha^2} \right)
\left( \ln r -\frac{H}{\lambda} \right)^2 \right].
\hspace{1cm}\mbox{for $\beta=0$}
\end{eqnarray} 

Figure 2(a) shows distributions $p(r)$ 
for case II and 
various $a$ with fixed values of 
$\lambda=1.0$, $b=1.0$, $I=0.1$, $\alpha=1.0$
and $\beta=0.0$ (multiplicative noise only).
With more decreasing $a$, a peak of $p(r)$
at $r \sim 0.1$ becomes sharper.
Fig. 2(c) shows
distributions $p(r)$
for case II and various $b$ with fixed values of
$\lambda=1.0$, $a=1.0$, $I=0.1$, $\alpha=1.0$
and $\beta=0.0$ (multiplicative noise only).
We note that a change in the $b$ value yields 
considerable changes in
shapes of $p(r)$.
Figs. 2(b) and 2(d) will be discussed shortly.

\subsubsection{Distribution of $T$}

When the temporal ISI $T$ is simply defined
by $T=1/r$, its distribution $\pi(T)$ 
is given by
\begin{eqnarray}
\pi(T)=p\left( \frac{1}{T} \right) \frac{1}{T^2}.
\end{eqnarray}
We get various distributions of $\pi(T)$ 
depending on functional forms of $F(x)$ and G(x).
For $F(x)=-\lambda x$, $G(x)=x$
and $\beta=0$, Eq. (26) yields
\begin{eqnarray}
\pi(T) \propto
T^{(2\lambda/ \alpha^2-1)}
\exp \left[ -\left(\frac{2H}{\alpha^2} \right) T \right], 
\end{eqnarray}
which expresses the gamma distribution [Eq. (2)]
\cite{Tuckwell88,Wilk00}.
For $F(x)=-\lambda x^2$, $G(x)=x$
and $\beta=0$,
Eq. (25) leads to
\begin{eqnarray}
\pi(T) \propto
T^{-1}
\exp\left[-\left( \frac{2H}{\alpha^2} \right)T
- \left( \frac{2\lambda}{\alpha^2} \right) \frac{1}{T} \right], 
\end{eqnarray}
which is similar to the inverse Gaussian distribution [Eq. (3)]
\cite{Gerstein64}.
For $F(x)=-\lambda \ln x$, $G(x)=x^{1/2}$ and $\beta=0$,
Eq. (31) yields
\begin{eqnarray}
\pi(T) \propto
T^{-3/2}
\exp\left[-\left( \frac{2\lambda}{\alpha^2} \right)
\left( \ln T +\frac{H}{\lambda} \right)^2 \right], 
\end{eqnarray}
which is similar to the log-normal distribution [Eq. (4)]
\cite{McKeegan02}.

Figs. 2(b) and 2(d) show $\pi(T)$
obtained from $p(r)$ shown in Figs. 2(a) and
2(c), respectively, by a change of variable with Eq. (32).
Fig. 2(b) shows that with more increasing $a$, 
the peak of $\pi(T)$ becomes sharper and moves left. 
We note in Fig. 2(d) that the form of $\pi(T)$ 
significantly varied by changing $b$ in $G(x)=x^b$. 

\subsubsection{Distribution of $R$}

When we consider the global variable $R(t)$ defined by
\begin{eqnarray}
R(t)&=&\frac{1}{N} \sum_{i} r_{i}(t),
\end{eqnarray}
the distribution $P(R,t)$ for $R$ is given by
\begin{equation}
P(R,t) = \int \cdots \int \Pi_i \:dr_{i} \:p(\{r_{i} \},t)
\: \delta\Bigl(R-\frac{1}{N}\sum_{j} r_{j}\Bigr).
\end{equation}

Analytic expressions of $P(R)$ are obtainable only for
limited cases.

\noindent
(a) For $\beta \neq 0$ and $ \alpha=0$, $P(R)$ is given by
\begin{eqnarray}
P(R) &\propto& 
\exp\left[ -\left( \frac{\lambda N}{\beta^2} \right)
\left( R- \frac{H}{\lambda} \right)^2 \right], 
\end{eqnarray}
where $H=H(I)$.

\noindent
(b) For $H=0$, we get \cite{Hasegawa07a}
\begin{equation}
P(R)=\frac{1}{2 \pi} \int_{-\infty}^{\infty}\: dk
\;e^{i k R}\:\Phi(k),
\end{equation}
with
\begin{equation}
\Phi(k)=\left[\phi\left( \frac{k}{N} \right) \right]^N,
\end{equation}
where $\phi(k)$ is the characteristic function
for $p(r)$ given by \cite{Abe00}
\begin{eqnarray}
\phi(k)&=& \int_{-\infty}^{\infty} \;
e^{-i k r}\:p(r) dr, \\
&=& 2^{1-\nu}\frac{(\lambda' \mid k \mid )^{\nu}}{\Gamma(\nu)}
K_{\nu}(\lambda' \mid k \mid),
\end{eqnarray}
with
\begin{eqnarray}
\nu&=& \frac{\lambda}{\alpha^2}, \\
\lambda'&=& \frac{\beta}{\alpha}, 
\end{eqnarray}
$K_{\nu}(x)$ expressing the modified Bessel
function.

Some numerical examples of $P(R)$ are
plotted in Figs. 3, 4 and 5.
Figures 3(a) and 3(b) show $P(R)$ for $\alpha=0.0$ 
and $\alpha=0.5$, respectively, when $N$ is changed.
For $\alpha=0.0$, $P(R)$ is the Gaussian distribution whose
width is narrowed by a factor of $1/\sqrt{N}$
with increasing $N$.
In contrast, $P(R)$ for $\alpha=0.5$ is 
non-Gaussian, whose shape seems to approach
a Gaussian for increasing $N$. These are consistent
with the central-limit theorem.

Effects of an external input $I$ on $p(r)$
and $P(R)$ are examined in Figs. 4(a) and 4(b). 
Figure 4(a) shows that in the case of $\alpha=0.0$
(additive noise only), $p(r)$ and $P(R)$ are simply
shifted by a change in $I$.
This is not the case for $\alpha \neq 0.0$,
for which $p(r)$ and $P(R)$ are shifted and 
widened with increasing $I$, as shown in Fig. 4(b).

Figures 5(a) and 5(b) show effects of the coupling $w$
on $p(r)$ and $P(R)$.
For $\alpha=0.0$, $p(r)$ and $P(R)$ are
changed only slightly with increasing $w$.
On the contrary, for $\alpha = 0.5$,
an introduction of the coupling
significantly modifies $p(r)$ and $P(R)$
as shown in Fig. 5(b).

\subsection{Dynamical properties}

\subsubsection{Augmented Moment Method (AMM)}

Next we will discuss the dynamical properties
of the rate model by using the AMM 
\cite{Hasegawa03a,Hasegawa06a,Hasegawa07a}.
By employing the FPE, we obtain
equations of motion 
for moments: $\langle r_{i}  \rangle$,
$\langle r_{i} \:r_{j} \rangle$, and
$\langle R^2 \rangle$ where $R=(1/N) \sum_i r_i$.
Then we get equations of motion for the three
quantities of $\mu$, $\gamma$ and $\rho$ defined by
\cite{Hasegawa03a,Hasegawa06a,Hasegawa07a}
\begin{eqnarray}
\mu &=& \langle R \rangle 
= \frac{1}{N} \sum_i \langle r_{i} \rangle, \\
\gamma &=& \frac{1}{N} \sum_i \langle (r_{i}-\mu)^2 \rangle, \\
\rho &=& \langle (R-\mu)^2 \rangle,
\end{eqnarray}
where $\mu$ expresses the mean, $\gamma$ the averaged
fluctuations in local variables ($r_{i}$) 
and $\rho$ fluctuations
in the global variable ($R$).
We get (for details see \cite{Hasegawa06a,Hasegawa07a})
\begin{eqnarray}
\frac{d \mu}{dt}&=& f_{0}+f_{2}\gamma + h_{0} 
+\left( \frac{\phi \: \alpha^2}{2}\right)
[g_{0}g_{1}+3(g_{1}g_{2}+g_{0}g_{3})\gamma], \\
\frac{d \gamma}{dt} &=& 2f_{1} \gamma
+ 2h_{1} \left( \frac{w N}{Z}\right) 
\left(\rho-\frac{\gamma}{N} \right) \nonumber \\
&+&(\phi+1) (g_{1}^2+2 g_{0}g_{2})\alpha^2\gamma
+ \alpha^2 g_{0}^2+\beta^2, \\
\frac{d \rho}{dt} &=& 2 f_{1} \rho 
+ 2 h_{1} w \rho
+ (\phi+1)
(g_{1}^2+2 g_{0}g_{2})\:\alpha^2 \:\rho 
+ \frac{1}{N}(\alpha^2 g_{0}^2  + \beta^2),
\end{eqnarray}
where $f_{\ell}=(1/\ell !)
(\partial^{\ell} F(\mu)/\partial x^{\ell})$,
$g_{\ell}=(1/\ell !)
(\partial^{\ell} G(\mu)/\partial x^{\ell})$, 
$h_{\ell}=(1/\ell !) 
(\partial^{\ell} H(u)/\partial u^{\ell})$ and
$u=w \mu +I$. 
Original $N$-dimensional stochastic DEs given by
Eqs. (5) and (6) are
transformed to the three-dimensional deterministic
DEs given by Eqs. (48)-(50).

When we adopt 
\begin{eqnarray}
F(x)&=&-\lambda x,  \\
G(x)&=& x,
\end{eqnarray}
Eqs. (48)-(50) are expressed 
in the Stratonovich representation ($\phi=1$) by
\begin{eqnarray}
\frac{d \mu}{dt}&=&-\lambda \mu + h_0
+ \frac{\alpha^2 \mu}{2}, \\
\frac{d \gamma}{dt} &=& -2 \lambda \gamma 
+ \frac{2 h_1 w N}{Z}\left( \rho-\frac{\gamma}{N} \right) 
+ 2 \alpha^2 \gamma + \alpha^2 \mu^2 + \beta^2, \\
\frac{d \rho}{dt} &=& -2\lambda \rho +2 h_1 w \rho
+ 2 \alpha^2 \rho
+ \frac{\alpha^2 \mu^2}{N}  + \frac{\beta^2}{N},
\end{eqnarray}
where $h_0=u/\sqrt{u^2+1}$, 
$h_1=1/(u^2+1)^{3/2}$ and $h_2=-(3 u/2)/(u^2+1)^{5/2}$
with $u=w \mu +I$.

Before discussing the dynamical properties,
we study the stationary properties of Eqs. (53)-(55).
We get the stationary solution given by
\begin{eqnarray}
\mu &=& \frac{h_0}{(\lambda-\alpha^2/2)}, \\
\gamma &=& \frac{(\alpha^2 \mu^2+\beta^2)}{2(\lambda-\alpha^2+w h_1/Z)}
\left[ 1+\frac{wh_1}{Z(\lambda-\alpha^2-w h_1)} \right], \\
\rho &=& \frac{(\alpha^2 \mu^2+\beta^2)}{2 N (\lambda-\alpha^2-w h_1)},
\end{eqnarray}
where Eq. (56) expresses the fifth-order algebraic equation of $\mu$.
The stability of Eqs. (53)-(55) around the stationary solution
may be shown by calculating eigenvalues
of their Jacobian matrix, although actual calculations
are tedious.

Figure 6 shows
the $N$ dependences of 
$\gamma$ and $\rho$ in the stationary state
for four sets of parameters:
$(\alpha, \beta, w)=(0.0, 0.1, 0.0)$ (solid curves),
(0.5, 0.1, 0.0) (dashed curves),
(0.0, 0.1, 0.5) (chain curves) and
(0.5, 0.1, 0.5) (double-chain curves),
with $\beta=0.1$, $\lambda=1.0$ and $I=0.1$.
For all the cases, $\rho$ is proportional to
$N^{-1}$, which is easily seen in Eq. (58).
In contrast, $\gamma$ shows a weak $N$ dependence
for a small $N$ ($< 10$).  
It is noted that $\sqrt{\gamma}$ and $\sqrt{\rho}$ approximately 
express the widths of $p(r)$ and $P(R)$, respectively.
The $N$-dependence of $\rho$
in Fig. 6 is consistent with the result shown in Figs. 3(a) and 3(b), 
and with the central-limit theorem.
\subsubsection{Response to pulse inputs}

We have studied the dynamical properties of the rate
model, by applying a pulse input of $I=I(t)$ given by
\begin{equation}
I(t)= A \:\Theta(t-t_{1}) \Theta(t_2-t)+I^{(b)},
\end{equation}
with $A=0.5$, $t_1=40$, $t_2=50$ and
$I^{(b)}=0.1$ which expresses the background input.
Figs. 7(a), 7(b) and 7(c) show the time dependences of
$\mu$, $\gamma$ and $\rho$ 
for $F(x)=-\lambda x$ and $G(x)=x$ 
when the input pulse
$I(t)$ given by Eq. (59) is applied \cite{Note1}:
solid and dashed curves show the results of the AMM and 
DS averaged over 1000 trials, respectively,
with $\alpha=0.5$, $\beta=1.0$, $w=0.5$ and $N=10$.
Figs. 7(b) and 7(c) show that
an applied input pulse induces changes in
$\gamma$ and $\rho$.
This may be understood from 
$2 \alpha^2$ terms in Eqs. (54) and (55).
The results of the AMM shown by solid curves in Figs. 7(a)-(c)
are in good agreement with DS results shown by dashed curves.
Figure 7(d) will be discussed shortly.

It is possible to discuss the synchrony in a neuronal ensemble
with the use of $\gamma$ and $\rho$ 
defined by Eqs. (46) and (47) \cite{Hasegawa03a}.
In order to quantitatively discuss the
synchronization, we first consider the quantity given by
\begin{equation}
P(t)=\frac{1}{N^2} \sum_{i j}<[r_{i}(t)-r_{j}(t)]^2>
=2 [\gamma(t)-\rho(t)].
\end{equation}
When all neurons are firing with the same rate
(the completely synchronous state),
we get $r_{i}(t)=R(t)$ for all $i$, and 
then $P(t)=0$ in Eq. (60).
On the contrary, we get 
$P(t)=2(1-1/N)\gamma \equiv P_{0}(t)$
in the asynchronous state
where $\rho=\gamma/N$ \cite{Hasegawa03a,Hasegawa07a}.
We may define the synchronization ratio
given by \cite{Hasegawa03a}
\begin{equation}
S(t) \equiv 1-\frac{P(t)}{P_{0}(t)}
= \left( \frac{N \rho(t)/\gamma(t)-1}{N-1} \right),
\end{equation}
which is 0 and 1 for completely asynchronous ($P=P_{0}$)  
and synchronous states ($P=0$), respectively.
Figure 7(d) shows
the synchronization ratio $S(t)$ for $\gamma(t)$ and $\rho(t)$
plotted in Figs. 7(b) and 7(c), respectively,
with $\alpha=0.5$, $\beta=1.0$, $w=0.5$ and $N=10$.
The synchronization ratio at $t < 40$ and $t > 60$
is 0.15, but it is decreased to 0.03 
at $40 < t < 50$ by an applied pulse.
This is because by an applied pulse,
$\gamma$ is more increased than $\rho$ and
the ratio of $\rho/\gamma$ is reduced.
The synchronization ratio vanishes for $w=0$,
and it is increased with increasing the coupling
strength \cite{Hasegawa03a,Hasegawa07a}.

Next we show some results for different indices of $a$ and $b$
in $F(x)=-\lambda x^a$ and $G(x)=x^b$.
Fig. 8(a) shows the time dependence of $\mu$ for
$(a, b)=(1, 1)$ (solid curve) and
$(a, b)=(2, 1)$ (dashed curve) 
with $\alpha=0.0$, $\beta=0.1$, $w=0.0$ and $N=10$.
The saturated magnitude of $\mu$ for $\alpha=0.5$
is larger than that for $\alpha=0.0$. 
Solid and dashed curves in Fig. 8(b) show $\mu$
for $(a,b)=(1,1)$ and (1,0.5), respectively,
with $\alpha=0.5$, $\beta=0.001$, $N=10$ and $w=0.0$.
Both results show similar responses to an applied
pulse although $\mu$ for a background input of $I^{(b)}=0.1$
for $(a, b)=(1,0.5)$ is a little larger than that
for $(a, b)=(1,1)$.

\subsubsection{Response to sinusoidal inputs}

We have applied also a sinusoidal input given by
\begin{equation}
I(t)= A \: \left[1-\cos \left( \frac{2 \pi t}{T_p} \right)\right]
+I^{(b)},
\end{equation}
for $F(x)=-\lambda \:x$ and $G(x)=x$
with $\lambda=1.0$, $A=0.5$, $I^{(b)}=0.1$, and $T_p=10$ and 20.
Time dependences of
$\mu$ for $T_p=20$ and $T_p=10$
are plotted in Figs. 9(a) and 9(b), respectively,
with $\alpha=0.5$, $\beta=1.0$, $w=0.0$ and $N=10$.
AMM results of $\mu(t)$ shown by solid curves in Figs. 9(a) and (b)
are indistinguishable from DS results (with 100 trials)
shown by dashed curves \cite{Note1},
chain curves denoting sinusoidal input $I(t)$. 
As the period of $T_p$ becomes shorter, the magnitude of $\mu$
becomes smaller. 
The delay time of $\mu(t)$ against an input $I(t)$
is about $\tau_d \sim 1.0$ ($= 1/\lambda$) for both $T_p=10$
and $T_p=20$.


\section{Discussion}
We may calculate variabilities  
of $r$ and $T$,
by using their distributions of $p(r)$ and $\Pi(T)$, which
have been obtained in Sec. 2.
For example, in the case of $F(x)=-\lambda x$ and
$G(x)=x$, the distribution of
$p(r)$ for $\beta=0.0$, $w=0.0$ and $H=I$
given by Eq. (26) leads to
\begin{eqnarray}
\langle r \rangle &=& \frac{I}{(\lambda-\alpha^2/2)}, \\
\langle \delta r^2 \rangle
&=& \langle (r-\langle r \rangle )^2 \rangle
= \frac{I^2 \alpha^2}{2 (\lambda-\alpha^2/2)^2(\lambda-\alpha^2)}, \\
\overline{c}_v &\equiv& 
\frac{\sqrt{\langle \delta r^2 \rangle}}{\langle r \rangle}
= \frac{\alpha}{\sqrt{2 (\lambda-\alpha^2)}}.
\end{eqnarray}
The relevant gamma distribution for ISI, $\Pi(T)$, given by Eq. (33) yields
\begin{eqnarray}
\langle T \rangle &=& \frac{\lambda}{I}, \\
\langle \delta T^2 \rangle &=&
\langle (T-\langle T \rangle )^2 \rangle
= \frac{\lambda \alpha^2}{2 I^2}, \\
c_v&\equiv& \frac{\sqrt{\langle \delta T^2 \rangle}}{\langle T \rangle}
= \frac{\alpha}{\sqrt{2 \lambda}}.
\end{eqnarray}
Equations (65) and (68) show that both $\overline{c}_v$ and $c_v$
are increased with increasing the magnitude ($\alpha$) 
of multiplicative noise.

It has been reported that 
spike train variability seems to correlated with location
in the processing hierarchy \cite{Harris05}.
A large value of $c_v$ is observed 
in hippocampus ($c_v \sim 3$) \cite{Fenton98} 
whereas $c_v$ is small in  cortical neurons  ($c_v \sim 0.5-1.0$) and
motor neurons ($c_v \sim 0.1$) \cite{Softky92}\cite{Calvin68}.
In order to explain the observed large $c_v$, several hypotheses
have been proposed: (1) a balance between excitatory and
inhibitory inputs \cite{Shadlen94}\cite{Troyer98}, 
(2) correlated fluctuations in recurrent 
networks \cite{Usher94},
(3) the active dendrite conductance \cite{Softky95},
and (4) a slowly decreasing tail of input ISI of
$T^{- d}$ ($d > 0$) at large $T$ \cite{Feng98}.
Our calculation shows that
multiplicative noise may be 
an alternative origin 
(or one of origins) of the observed
large variability.
We note that the variability of $r$ is given by
$\overline{c}_v=\sqrt{\gamma}/\mu$ in the AMM ({\it e.g.}
Eq. (65) agrees with $\sqrt{\gamma}/\mu$ for
$\mu$ and $\gamma$
given by Eqs. (56) and (57), respectively, with $w=\beta=0$).
It would be interesting to make a more detailed study of
the variability 
for general $F(x)$ and $G(x)$ as discussed in Sec. 2.

We have proposed the generalized rate-code model 
given by Eqs. (5) and (6),
in which the relaxation process is given by
a single $F(x)$.
Instead, when the relaxation process consists of two terms:
\begin{equation}
F(x) \rightarrow c_1 F_1(x)+c_2 F_2(x),
\end{equation}
with $c_1+c_2=1$,
the distribution becomes
\begin{equation}
p(r)=[p_1(r)]^{c_1}\:[p_2(r)]^{c_2},
\end{equation}
where $p_k(r)$ ($k=1, 2$) denotes the distribution only
with $F(x)=F_1(x)$ or $F(x)=F_1(x)$.
In contrast, when multiplicative noise
arises from two independent origins:
\begin{equation}
\alpha x \eta(t) 
\rightarrow c_1 \alpha_1 x \eta_1(t)
+c_2 \alpha_2 x \eta_2(t),
\end{equation}
the distribution for $\beta=H=0$ becomes 
\begin{equation}
p(r) \propto r^{-[2 \lambda/(c_1 \alpha_1^2+c_2 \alpha_2^2)+1]}.
\end{equation}
Similarly, when additive noise
arises from two independent origins:
\begin{equation}
\beta \xi(t) 
\rightarrow c_1 \beta_1 \xi_1(t)+c_2 \beta_2 \xi_2(t),
\end{equation}
the distribution for $\alpha=H=0$ becomes 
\begin{equation}
p(r) \propto {\rm e}^{-\lambda/(c_1 \alpha_1^2+c_2 \alpha_2^2)}.
\end{equation}
Equations (70), (72) and (74) are quite different from the
form given by
\begin{equation}
p(r)=c_1 p_1(r)+c_2 p_2(r),
\end{equation}
which has been conventionally adopted for a fitting of the
theoretical distribution to that obtained by experiments
\cite{Gerstein64}-\cite{Vog05}.

\section{Conclusion}
We have proposed the generalized rate-code model
[Eqs. (5) and (6)],
whose
properties have been discussed
by using the FPE and the AMM.
The proposed rate model is a phenomenological one and
has no biological basis.
As discussed in Sec. 1, the conventional 
rate model given by Eq. (1)
may be obtainable from a spiking neuron model
when we adopt appropriate approximations to DEs
derived by various approaches such as
the population-density method 
\cite{Wilson72}-\cite{Eggert00}
and others \cite{Amit91}-\cite{Shriki03}.
It would be interesting to derive our rate model
given by Eqs. (5) and (6), starting from a spiking neuron model.
The proposed generalized rate model is useful
in discussing stationary and dynamical
properties of neuronal ensembles.
Indeed, our rate model has an interesting property,
yielding various types of
stationary non-Gaussian distributions
such as gamma, inverse-Gaussian 
and log-normal distributions,
which have been experimentally observed
\cite{Gerstein64}-\cite{Vog05}.
It is well known that the Langevin-type model given by Eq. (1) 
cannot properly describe fast neuronal dynamics
at the characteristic times $\tau_c$ shorter 
than $\tau$ ($\equiv 1/\lambda $).
This is, however, not a fatal defect because
we may evaded it, by adopting an appropriate $\tau$ value
of $\tau < \:\tau_c/10$ for a given neuronal ensemble with $\tau_c$.
Actually, the dynamical properties of an ensemble consisting
of excitatory and inhibitory neurons has been successfully discussed
with the use of the Langevin-type Wilson-Cowan model 
\cite{Wilson72}\cite{Amari72} 
(for recent papers using the Wilson-Cowan model, see \cite{WCmodel}:
related references therein).
One of the disadvantages of the AMM is that its applicability
is limited to the case of weak noise because it neglects
contributions from higher moments.

On the contrary, the AMM has following advantages:

\noindent
(i) the dynamical properties of an $N$-unit neuronal ensemble
may be easily studied by solving three-dimensional ordinary DEs 
[Eqs. (48)-(50)],
in which three quantities of $\mu$, $\gamma$ and $\rho$
have clear physical meanings,

\noindent
(ii) analytic expressions for
DEs provide us with physical insight without numerical calculations
({\it e.g.} the $N$ dependence of $\rho$ follows the central-limit 
theorem [Eq. (58)],
and

\noindent
(iii) the synchronization of the ensemble may be discussed [Eq. (61)].

\noindent
As for the item (i), 
note that we have to solve
the $N$-dimensional stochastic Langevin equations 
in DS, and
the $(2N+1)$-dimensional partial DEs in the FPE.
Then the AMM calculation is very much faster than DS:
for example, for the calculation shown in Fig. 9(a), 
the ratio of the computation time of the AMM to that of DS becomes
$t_{AMM}/t_{DS} \sim $1/$30 \:000$ \cite{Note2}.
We hope that the proposed rate model may be adopted
for a wide class of study on neuronal ensembles
described by the Wilson-Cowan-type model \cite{Hasegawa07}.

\section*{Acknowledgements}
This work is partly supported by
a Grant-in-Aid for Scientific Research from the Japanese 
Ministry of Education, Culture, Sports, Science and Technology.




\newpage

\begin{figure}
\caption{
(a) Distributions $p(r)$ of the (local) firing rate $r$
for various $\alpha$ with $\lambda=1.0$, 
$\beta=0.1$, $I=0.1$ and $w=0.0$,
(b) $p(r)$
for various $\beta$ with $\lambda=1.0$, 
$\alpha=1.0$, $I=0.1$ and $w=0.0$, and
(c) $p(r)$
for various $I$ with $\lambda=1.0$, 
$\alpha=0.5$, $\beta=0.1$ and $w=0.0$.
}
\label{fig1}
\end{figure}

\begin{figure}
\caption{
(a) Distributions $p(r)$ of the (local) firing rate $r$
and (b) $\pi(T)$ of the ISI $T$
for $a=0.8$ (chain curves), $a=1.0$ (solid curves),
$a=1.5$ (dotted curves) and $a=2.0$ (dashed curves)
with $\lambda=1.0$, $b=1.0$, $I=0.1$,
$\alpha=1.0$ and $\beta=0.0$ (multiplicative noise only).
(c) $p(r)$ of the (local) firing rate $r$
and (d) $\pi(T)$ of the ISI $T$
for $b=0.5$ (dashed curves),
$b=1.0$ (solid curves), $b=1.5$ (dotted curves)
and $b=2.0$ (chain curves) 
with $\lambda=1.0$, $a=1.0$, $I=0.1$, 
$\alpha=1.0$ and  $\beta=0.0$ (multiplicative noise only):
results for $b=1.5$ and $b=2$ should be multiplied 
by factors of 2 and 5, respectively.
}
\label{fig2}
\end{figure}

\begin{figure}
\caption{
Distributions $P(R)$ of the (global) firing rate $R$
for (a) $\alpha=0.0$
and (b) $\alpha=0.5$, with $N=1$, 10 and 100:
$\lambda=1.0$, $\beta=0.1$, $w=0.0$
and $I=0.1$.
}
\label{fig3}
\end{figure}

\begin{figure}
\caption{
Distributions $p(r)$ (dashed curves)
and $P(R)$ (solid curves)
for (a) $\alpha=0.0$ and (b) $\alpha=0.5$ 
with $I=0.1$ and $I=0.2$:
$N=10$, $\lambda=1.0$, $\beta=0.1$ and $w=0.0$.
}
\label{fig4}
\end{figure}

\begin{figure}
\caption{
Distributions $p(r)$ (dashed curves)
and $P(R)$ (solid curves)
for (a) $\alpha=0.0$ and (b) $\alpha=0.5$
with $w=0.0$ and $w=0.5$:
$N=10$, $\lambda=1.0$, $\beta=0.1$ and $I=0.1$.
}
\label{fig5}
\end{figure}

\begin{figure}
\caption{
The $N$ dependence of 
$\gamma$ and $\rho$ in the stationary states 
for four sets of parameters:
$(\alpha, \beta, w)=(0.0, 0.1, 0.0)$ (solid curves),
$(0.5, 0.1, 0.0)$ (dashed curves),
$(0.0, 0.1, 0.5)$ (chain curves) and
$(0.5, 0.1, 0.5)$ (double-chain curves):
$\lambda=1.0$, $N=10$ and $I=0.1$.
}
\label{fig6}
\end{figure}

\begin{figure}
\caption{
(Color online)
Time courses of 
(a) $\mu(t)$, (b) $\gamma(t)$, (c) $\rho(t)$
and (d) $S(t)$
for a pulse input $I(t)$ given by Eq. (59)
with $\lambda=1.0$,
$\alpha=0.5$, $\beta=0.1$, $N=10$ and $w=0.5$,
solid and chain curves 
denoting results  of AMM and dashed curves expressing
those of DS result with 1000 trials.
}
\label{fig7}
\end{figure}

\begin{figure}
\caption{
(a) Response of $\mu(t)$ to input pulse $I(t)$ 
given by Eq. (59)
for $(a,b)=(1, 1)$ (solid curve) and 
$(a, b)=(2, 1)$ (dashed curve) 
with $\alpha=0.0$, $\beta=0.1$,
$N=10$ and $\lambda=1.0$.
(b) Response of $\mu(t)$ to input pulse $I(t)$
for $(a,b)=(1, 1)$ (solid curve) and 
$(a, b)=(1, 0.5)$ (dashed curve) 
with $\alpha=0.5$, $\beta=0.001$,
$N=10$, $\lambda=1.0$ and $w=0.0$.
}
\label{fig8}
\end{figure}

\begin{figure}
\caption{
(Color online)
Response of $\mu(t)$ 
to sinusoidal input $I(t)$ (chain curves)
given by Eq. (62)
for (a) $T_p=20$ and (b) $T_p=10$
with $A=0.5$, $\lambda=1.0$, $\alpha=0.5$, 
$\beta=0.1$, $w=0.0$ and $N=10$ 
($a=1$ and $b=1$), solid and chain curves
denoting $\mu(t)$ of AMM and dashed curves expressing
those of DS result with 100 trials.
}
\label{fig9}
\end{figure}

\end{document}